\documentclass[%
 reprint,
 amsmath,amssymb,
 aps, prl
]{revtex4-1}

\usepackage{graphicx}
\usepackage{dcolumn}
\usepackage{bm}

\newcommand{\be}{\begin{equation}}
\newcommand{\ee}{\end{equation}}
\newcommand{\bea}{\begin{eqnarray}}
\newcommand{\eea}{\end{eqnarray}}
\newcommand{\pd}{\partial}

\begin{document}

\preprint{APS/123-QED}

\title{Lossless Suppression and Enhancement of Soliton Self-Frequency Shifts} 

\author{Francisco R. Arteaga-Sierra$^{1}$}
 \email{f.arteaga-sierra@rochester.edu}
\author{Aku Antikainen$^{1,}$}%
\author{Govind P. Agrawal$^{1,2}$}
\affiliation{%
 $^1$The Institute of Optics, University of Rochester, Rochester, New York 14627 \\
   $^2$Laboratory for Laser Energetics, 250 East River Rd, Rochester, NY 14623
}%

\date{\today}

\begin{abstract}
Soliton self-frequency shifts (SSFS) can be suppressed in optical fibers through spectral recoil, but this process leads to losses through continuous transfer of energy to a dispersive wave. We demonstrate a novel way to alter the strength of SSFS in photonic crystal fibers through a frequency-dependent nonlinear parameter $\gamma(\omega)$. Our numerical simulations show both suppression and enhancement of SSFS depending on the sign of nonlinear slope $\pd \gamma/ \pd \omega$. A large enough positive value of this slope can lead to total suppression of SSFS, without spectral recoil and without energy transfer to a resonant dispersive wave. Numerical simulations are supported by mathematical predictions based on the moment method.
\begin{description}
\item[PACS numbers] 42.81.Dp, 42.65.Ky, 42.65.-k

\end{description}
\end{abstract}

\maketitle

\section{\label{sec1} Introduction}

The soliton self-frequency shift (SSFS) is one of the most important nonlinear processes involved in supercontinuum generation using optical fibers (see Refs.\ \cite{Dudley:06,Skryabin:10} for reviews on the topic). Characterized by a continuous downshift of the central frequency of sub-picosecond pulses \cite{Gordon:86, Mitschke:86}, the SSFS is in part responsible for effects such as the creation of dispersive waves (DWs) at new frequencies \cite{Akhmediev:95, Roy:11}, trapping of DWs by solitons \cite{Gorbach:07,Gorbach:07b}, formation of spectral cavities, and enhancement of spectral broadening \cite{Driben:13}, to mention only a few examples. Some studies have reported the enhancement of the SSFS using uniform and tapered photonic crystal fibers (PCFs) \cite{Judge:09,Pant:10,arteaga:14,antikainen:17}. On the other hand, suppression of the SSFS of ultrashort pulses has been also demonstrated in conventional fibers \cite{Skryabin:03,Biancalana:04} and in PCFs doped with silver nanoparticles \cite{Bose:16a,Bose:16b}. In these works, the suppression of the SSFS is accompanied with the transfer of energy from the propagating Raman soliton to a DW on the red side and soliton's subsequent recoil towards the blue side. In other words, the continuous compensation of SSFS happens through spectral recoil.

In this article, we study numerically the evolution of ultrashort pulses in PCFs whose nonlinear Kerr response is strongly frequency-dependent. Even though the nonlinear parameter $\gamma$ of PCFs is often treated as being constant or varying only slightly varying with frequency, $\gamma$ can become strongly frequency dependent in fibers doped with silver nanoparticles \cite{Driben:09,Driben:10} or waveguides employing quadratic nonlinear media \citep{Moses:06}. Here, we assume a linear variation of $\gamma$ with frequency, without focusing on the mechanism that leads to such variations. When the slope $\gamma_1 = \pd \gamma / \pd \omega < 0$, the nonlinearity increases with wavelength, causing enhanced SSFS for red-shifting solitons. Conversely, suppression of SSFS occurs for $\gamma_1>0$. For a sufficiently large value of $\gamma_1$, we show that SSFS can be totally suppressed. Unlike the other reported mechanisms of Raman-shift suppression \cite{Skryabin:03,Biancalana:04, Bose:16b}, SSFS suppression in our case occurs without the loss of energy to a DW at a new phase-matched wavelength. Instead, the suppression is caused by reshaping of the soliton induced by the frequency-dependent nonlinearity. Since this new process of SSFS suppression does not continuously transfer energy to a DW, the Raman soliton does not lose energy while propagating inside the fiber.

\section{Impact of $\gamma_1$ on Soliton Self-Frequency Shifts}

We use the well-known generalized nonlinear Schr\"odinger equation in the spectral domain~\cite{Agrawal:13},
\begin{align}\label{eq_GNLSE}
    \frac{\partial\tilde A}{\partial z} - i [\beta(\omega)-\beta(\omega_0)- \beta_1(\omega-\omega_0)] \tilde A = \hspace{1cm} \hspace{0.5cm} \\
    i\gamma(\omega) (1 - f_R){\hat{\cal F}} \left(|A|^2A \right) + \gamma_R(\omega) f_R\times\nonumber \hspace{1cm} \\ {\hat{\cal F}}\left( A \int^{\infty}_{-\infty} h_R(T') |A(z,T-T')|^2dT'\right), \hspace{0cm} \nonumber
\end{align}
where $\hat{\cal F}$ is the Fourier-transform operator, $\tilde A(z,\omega)=\hat{\cal F}[A(z,t)]$ is the Fourier transform of the pulse envelope $A(z,t)$, $\beta(\omega)$ is the propagation constant of the PCF mode, $\beta_1 = d\beta/d\omega$ is calculated at the carrier frequency $\omega_0=2\pi c /\lambda_0$ ($\lambda_0=1060$ nm, anomalous regime) of the pulse, and $T=t-\beta_1 z$ is the time measured in a frame moving at group velocity of input pulse. The nonlinear effects are included through the nonlinear parameter $\gamma(\omega)=\gamma_0+\gamma_1(\omega-\omega_0)$, where $\gamma_0=\gamma(\omega_0)$ and $\gamma_1=d\gamma/d\omega$ is the nonlinear slope (parameter leading to self-steepening). The Raman nonlinearity is taken to be that of silica: $\gamma_R(\omega)=\gamma_0+\gamma_{1R}(\omega-\omega_0)$ with $\gamma_{1R}= \gamma_0/\omega_0$. The Raman fraction is $f_R=0.18$ and for $h_R(t)$ we use the common Raman response function for silica \cite{Agrawal:13}.

\begin{figure}[tb!]
\includegraphics[width=\linewidth]{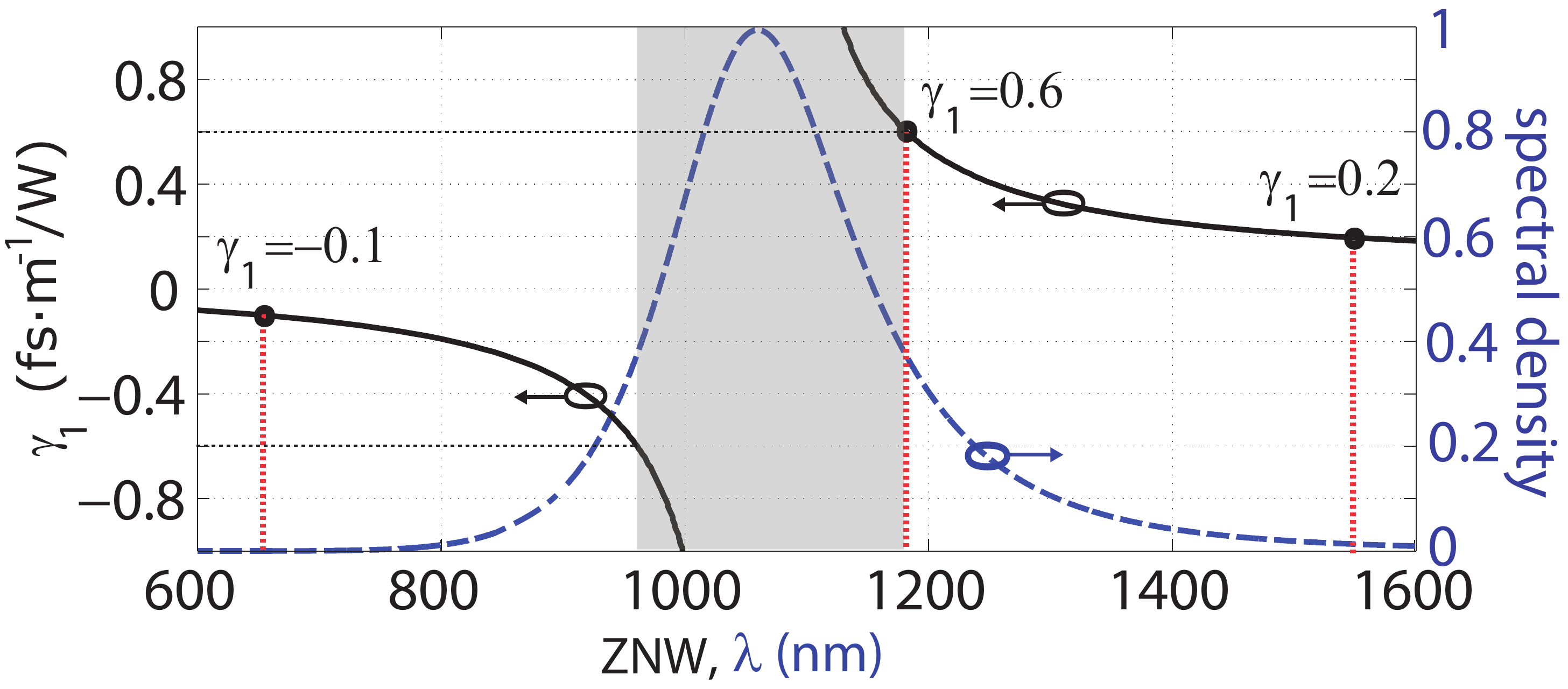}
\caption{(Color online) $\gamma_1$ versus ZNW (solid black), and spectrum of input pulses (dashed blue). The red vertical dashed lines mark the ZNW for three different values of $\gamma_1$. The grey region shows the range of ZNWs in which the moment method cannot be used to accurately describe the SSFS.}
\label{fig:ZNWVsgamma}
\end{figure}

\begin{figure}[tbp!]
\includegraphics[width=\linewidth]{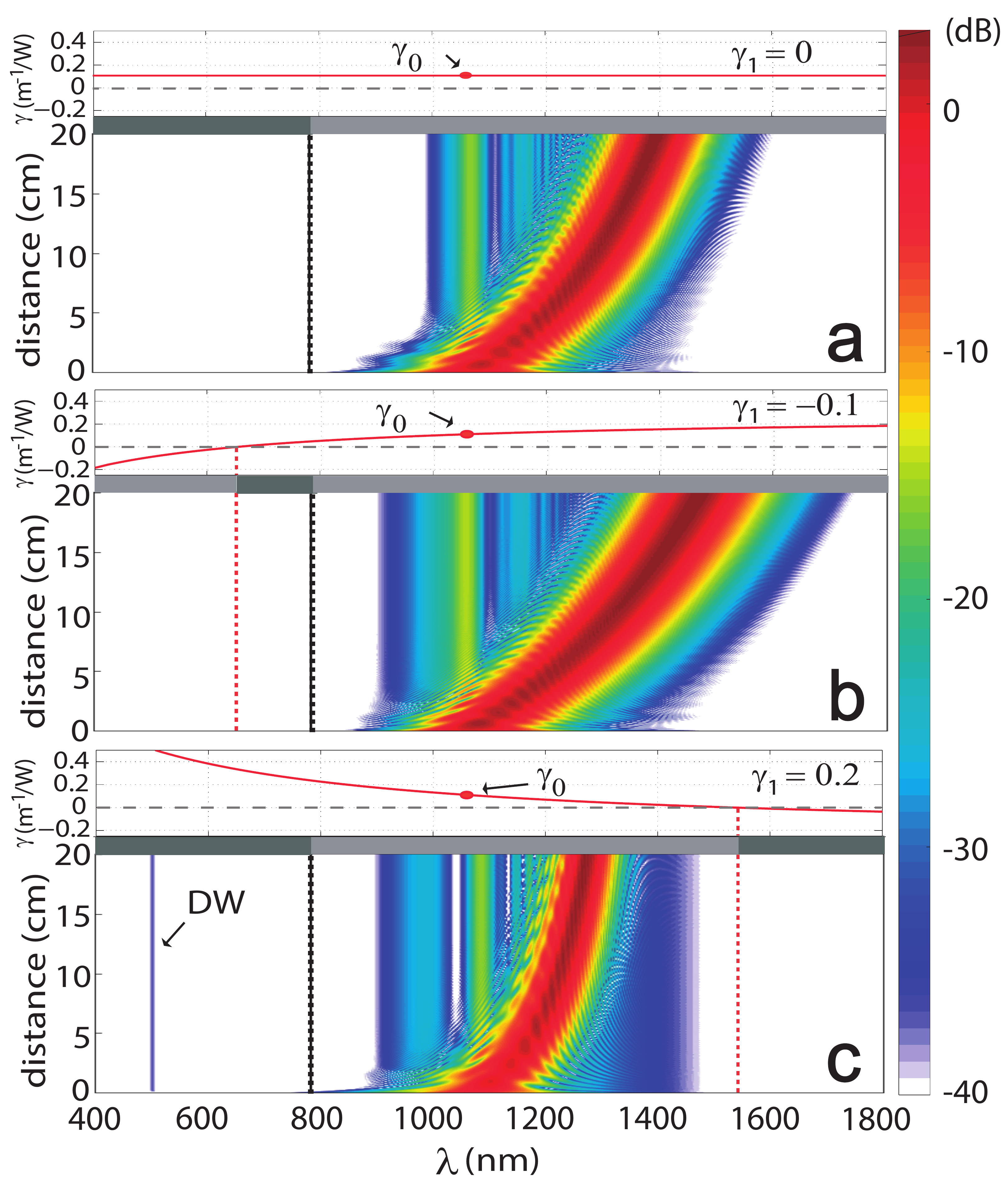}
\caption{(Color online) Spectral evolution of the 10-fs pulse in three $20$-cm-long PCFs with (a) $\gamma_1=0$, (b) $-0.1$, and (c) $0.2$ fs/W-m. The ZDW at $780$ nm is marked by a dashed black line, and the ZNWs are marked by red dashed lines. Top traces show $\gamma(\omega)$ and the grey bars show the regions where solitons can (light grey) or cannot (dark grey) form.}
\label{fig:Evos}
\end{figure}

Equation (\ref{eq_GNLSE}) is solved numerically using the fourth-order Runge--Kutta method for a 20-cm-long PCF with its zero-dispersion wavelength (ZDW) at $780$~nm. Our PCF is identical to the one used in Ref.~\cite{Ranka:00} with $\gamma_0=0.11$ W$^{-1}$/m. However, its nonlinear slope $\gamma_1$ is varied between $-0.6$ and $0.6$~fs/(W-m). Since $\gamma_0$ is acting as a pivot for varying $\gamma(\omega)$, the sign of $\gamma_1$ determines where the zero-nonlinearity wavelength (ZNW) is located relative to the input wavelength of 1060~nm. Figure \ref{fig:ZNWVsgamma} shows the correspondence between the ZNW and $\gamma_1$. Numerical simulations are performed for an ultrashort optical pulses launched such that it forms a fundamental soliton initially. More specifically, we solve Eq.\ (\ref{eq_GNLSE}) with the input $A(0,T)=\sqrt{P_0}$ sech$ (T/T_0)$ with $T_0 \approx 10/1.763$~fs (full width at half maximum 10~fs). The peak power of input pulse is chosen to be $P_0=14.3$~kW so that the input soliton order is $N=T_0\sqrt{\gamma_0 P_0/|\beta_2(\omega_0)|} = 1$. Figure \ref{fig:Evos} compares the spectral evolutions of the pulse in three PCFs. Part (a) shows the evolution in a fiber with constant nonlinearity ($\gamma_1=0$), whereas parts (b) and (c) correspond to PCFs with $\gamma_1=-0.1$ and $\gamma_1=0.2$, respectively. In each case, the wavelength dependence of $\gamma$ is shown on top, together with a grey bar marking the spectral regions in which solitons can (light grey) or cannot (dark grey) form. The formation of solitons requires the nonlinearity $\gamma$ and the group velocity dispersion $\beta_2$ to have opposite signs. 

The spectral evolution inside the constant-$\gamma$ fiber ($\gamma_1=0$) in Fig.\ \ref{fig:Evos}(a) shows the expected red shift reaching 340~nm within 20~cm of fiber. The central frequency of the soliton is $\lambda_s\approx1400$~nm at the output of the fiber. The fiber in Fig.~\ref{fig:Evos}(b) has $\gamma_1=-0.1$, with its ZNW located at $\approx655$~nm (see Fig.~\ref{fig:ZNWVsgamma}). The soliton's red shift is enhanced in this situation, and its central wavelength is close to $\lambda_s\approx 1470$~nm at the PCF output. In contrast, the fiber in Fig.~\ref{fig:Evos}(c) with $\gamma_1=0.2$ has its ZNW near 1530~nm, and soliton's red shift is considerably reduced such that $\lambda_s\approx1270$~nm dat the PCF output. Clearly, the SSFS is considerably suppressed compared to the $\gamma_1=0$ case shown in part (a). As expected on physical grounds, positive values of $\gamma_1$ lead to an enhanced red shift, while its negative values reduce it, and the soliton appears to be repulsed away from the ZNW of the fiber.

\section{Predictions of the Moment Method}

To explain large differences in SSFS magnitudes seen in Fig.~\ref{fig:Evos}, we can refer to the \textit{Gordon formula} obtained using a perturbation theory of solitons \cite{Gordon:86}. This formula shows a strong dependence of the SSFS on the temporal width of the pulse, which changes inside the fiber with propagation. Additionally, it has been demonstrated by using the moment method that the SSFS is also dependent on the frequency chirp that invariably occurs in our PCFs \cite{Santhanam:03}. A more detailed analysis of the SSFS has been developed by Chen \textit{et al.} \cite{Chen:10} for ultrashort pulses using an improved moment method. Following the same methodology, we obtain the following set of ordinary differential equations describing the evolution of various pulse parameters:
\begin{align}
    \label{E}  \frac{dE}{dz}   &= -\alpha E - \left(\frac{4\gamma_{1R}T_R}{15}\right)\frac{E^2}{T_0^3}, \\
    \label{T} \frac{dT_p}{dz}    &= \beta_2\Omega+\frac{\beta_3}{2} \left[\Omega^2+\left( 1+\frac{\pi^2 C^2}{4} \right)\frac{1}{3T_0^2}\right]+\frac{\gamma_1}{2T_0}E, \\
    \label{T0} \frac{dT_0}{dz} &= \frac{\left(\beta_2+\beta_3 \Omega \right) C}{T_0}+\left(\frac{4\gamma_{1R}T_R}{\pi^2}\right)\frac{E}{T_0^2}, 
\end{align}    

\begin{align}
    \frac{dC}{dz}    &= \left(\frac{4}{\pi^2}+C^2\right)
    \frac{\left(\beta_2+\beta_3  \Omega\right)}{T_0^2} \nonumber \\
    &+ \left[\frac{\left(150-4\pi^2\right)}{15\pi^2}\right]
    \frac{\gamma_{1R}T_RE C}{T_0^3} \label{C}  \\
    &+\frac{2}{\pi^2}\left[1+6f_R D(T_0)\right] \left(\gamma_0+\gamma_1\Omega\right)\frac{E}{T_0}, \nonumber
\end{align}
where $E$ is the energy of the pulse, $T_p$ and $T_0$ are its temporal position and width, and $C$ is the time-domain chirp parameter. The effective Raman parameter $T_R$ and the coefficient $D$ are functions that depend on the temporal width $T_0$ of the pulse (see \cite{Chen:10} for details). Since losses have been neglected in our simulations ($\alpha=0$), the pulse can only lose energy through intrapulse Raman scattering (second term in Eq.~\ref{E}). The SSFS, $\Omega = \omega_0 - \omega$, evolves according to
\begin{align}
	\label{Omega} \frac{d\Omega}{dz}=-\frac{4\left(\gamma_0+
    \gamma_{1R}\Omega\right) T_R E}{15T_0^3}+\frac{\gamma_1CE}{3T_0^3}.
\end{align}
This equation shows that the frequency shift $\Omega$(z) depends on not only on the pulse width $T_0$ but also on pulse energy $E$ and Chirp $C$, which themselves evolve with $z$ as indicated in Eqs.\ (\ref{E})--(\ref{Omega}). It is clear from Eq.\ (\ref{Omega}) that the first term on its right side is always negative (leading to a red shift) but the second term can be positive or negative depending on the signs of $\gamma_{1}$ and $C$. In particular, this term is positive for positive values of $\gamma_{1}$ and $C$ and thus can compensate for the negative first term, resulting in smaller red-shifts, as seen in Fig.~\ref{fig:Evos}(c). 

\begin{figure}[tpb!]
\centering
\includegraphics[width=8cm]{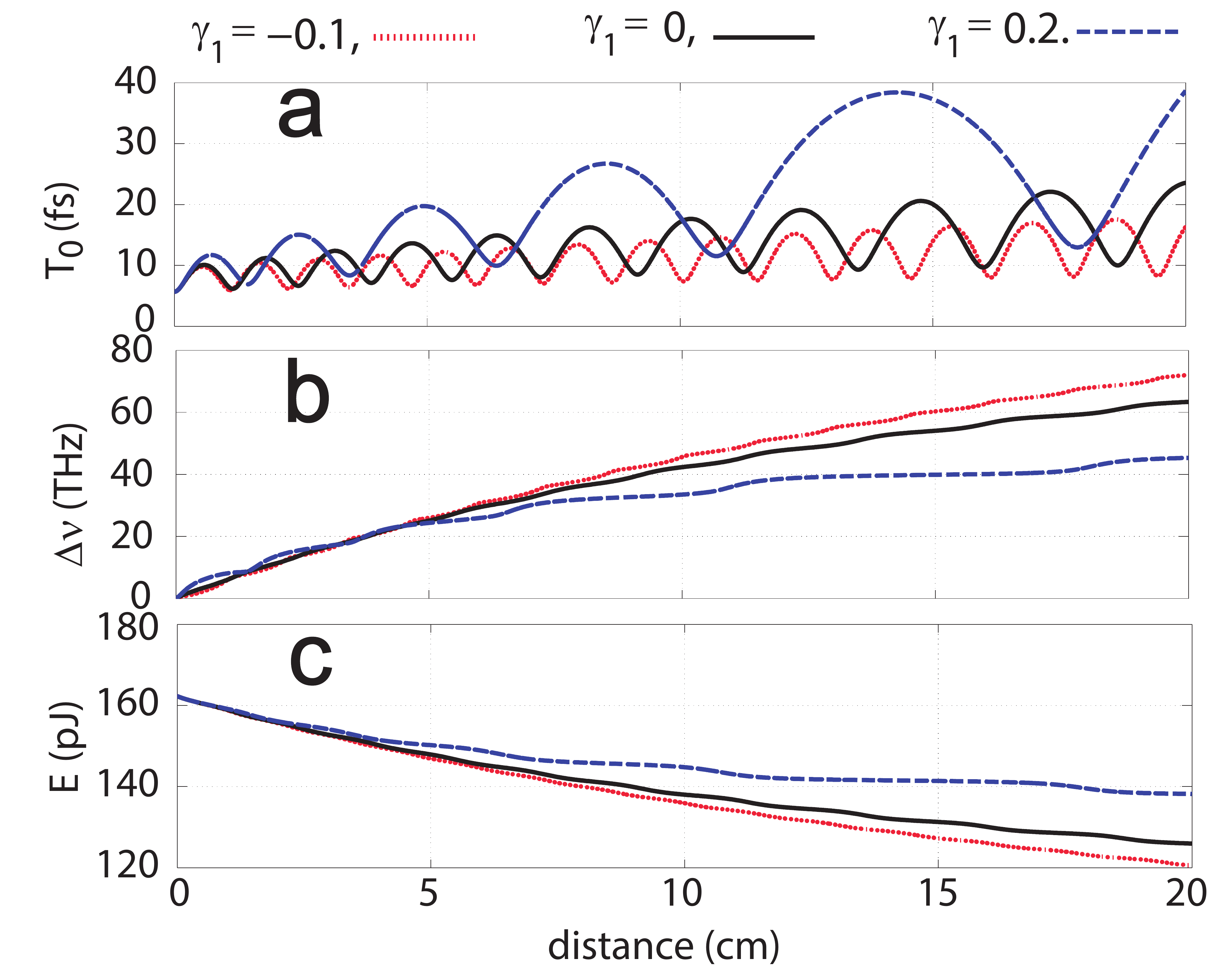}
\caption{(Color online) Moment-method predictions for the evolutions of (a) temporal width, (b) frequency shift, and (c) energy along the three PCFs used in Fig.\ \ref{fig:Evos}.}
\label{fig:shift}
\end{figure}

To compare the predictions of the moment method with the simulations, we solve Eqs.\ (\ref{E}--\ref{Omega}) using the fourth-order Runge--Kutta method for the three fibers used in Fig. \ref{fig:Evos}. Figure \ref{fig:shift} shows evolutions of the temporal width $T_0$, frequency shift $\Delta\nu=|\Omega|/2\pi$, and the energy $E$ for the three values of $\gamma_1$. The temporal width in part (a) exhibits an oscillatory behavior in all cases. The oscillations result from changes in the chirp parameter $C$ (not shown) that oscillates between positive and negative values causing compression of the pulse when $\beta_2C<0$. The duration $T_0$ of the pulse increases to compensate for the decreasing $\gamma$ at the soliton's shifting central frequency to maintain the soliton condition $N = T_0(z)\sqrt{\gamma(z) P_0(z)/|\beta_2(z)|} = 1$. Another interpretation is that the decreasing $\gamma$ increases the nonlinear length defined as $L_{NL}=1/\gamma(\omega)/P_0$, thus making the soliton more sensitive to pulse-spreading dispersive effects. The predictions of the SSFS in Fig.\ \ref{fig:shift}(b) are in reasonable agreement with the numerical results shown in Fig.~\ref{fig:Evos}. Small discrepancies can be attributed to the moment method not accounting for the emission of DWs. Also, the spectral recoil caused by the emission of a blue DW can contribute to the mismatch if the resonant radiation has enough intensity to undergo a significant initial red shift. However, the total error in the prediction of the moment-method SSFS is less than 10\% in all cases shown in Fig.\ \ref{fig:Evos}. Raman scattering causes an optical pulse to transfer some energy to the medium. The magnitude of the loss through intrapulse Raman scattering (second term in Eq.~\ref{E}) is smaller for larger values of $T_0$. Therefore, longer pulses experience less SSFS and the increase in pulse duration helps combat the red shift, as seen in \ref{fig:shift}(c). In the absence of fiber losses, the conservation of the number of photons, $n_{ph} = E(z)/\mathbf{h}\nu_0(z)$, supports the validity of Eqs.\ (\ref{E})--(\ref{Omega}). Physically speaking, the initial red shift and changes in the local nonlinearity cause the soliton to increase its temporal width in order to preserve its soliton status by maintaining $N=1$. The preservation of the soliton status also means that the soliton cannot enter the non-solitonic regime where $\beta_2$ and $\gamma$ have the same signs. When this regime is on the red side of the soliton, the red shift has to be reduced considerably for the soliton to avoid crossing over.

\section{Total suppression of SSFS}

\begin{figure}[tbp]
\includegraphics[width=8cm]{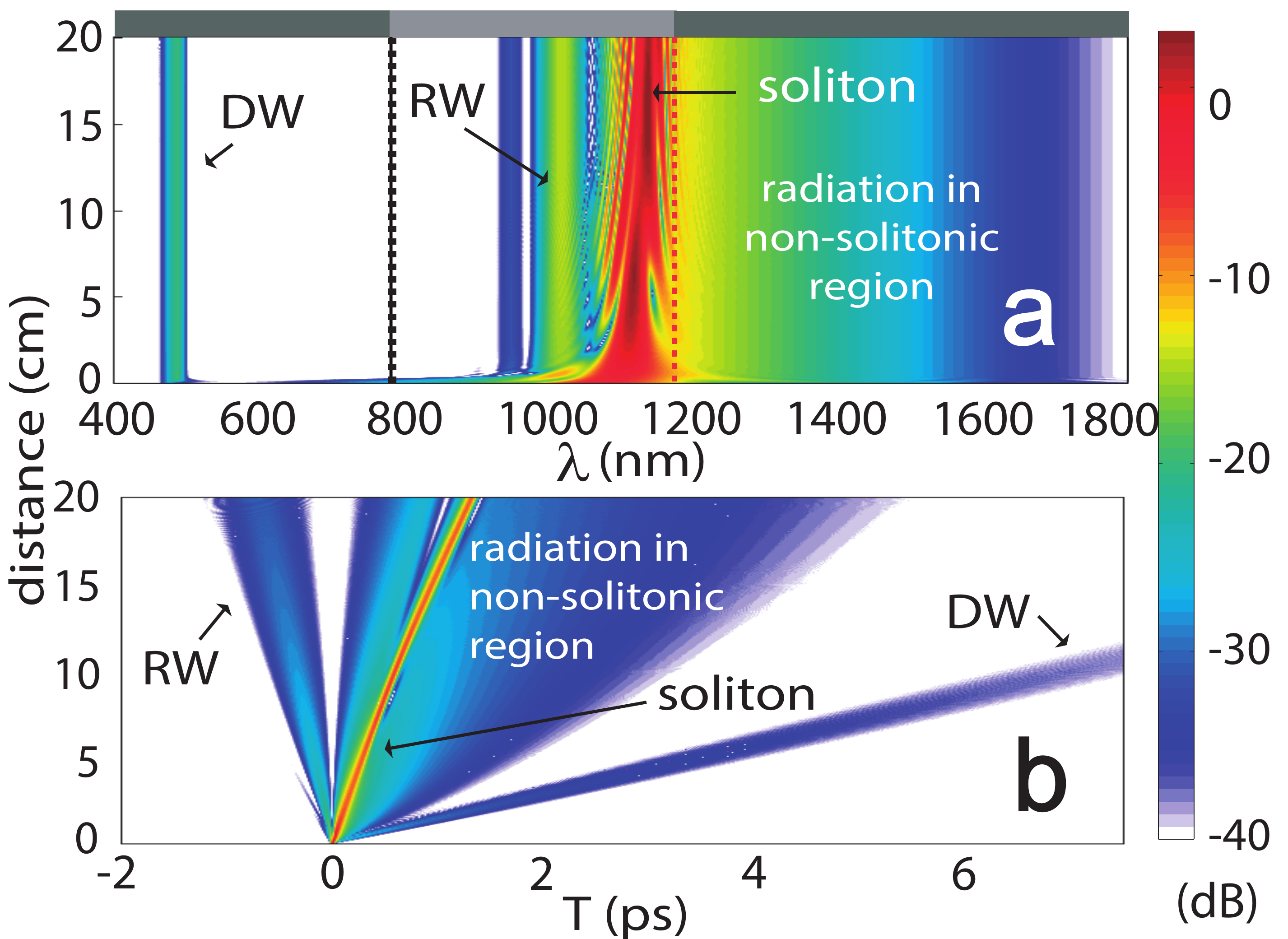}
\caption{(Color online) (a) Spectral and (b) temporal evolutions of the fundamental soliton under the conditions of Figs. \ref{fig:Evos}, except that $\gamma_1=0.6$. The ZDW is marked by a black line, and the red dashed line show the ZNW.}.
\label{fig:Evos6}
\end{figure}

It becomes natural to ask whether engineering the fiber nonlinearity through changing $\gamma_1$ can lead to total suppression of the SSFS\@. Since $\gamma_1 > 0$ hinders SSFS, its large enough positive value could be expected to lead to better SSFS suppression. Figure \ref{fig:Evos6} shows the spectral and temporal evolution of the 10-fs pulse under the conditions of Fig.~\ref{fig:Evos} except that $\gamma_1=0.6$.
The spectral evolution in part (a) shows that the spectrum shifts by 60~nm within the first few centimeters of the PCF and then stops shifting, indicating a nearly total suppression of the Raman shift after $10$ cm of propagation. More specifically, the Raman soliton stays at $\lambda_s=1120$~nm after $10$~cm of such that the entire pulse spectrum is to the left of the ZNW located near 1180~nm. Also note that the emitted DW has more energy compared to that in Fig.~\ref{fig:Evos}. The suppression of SSFS keeps the soliton confined in a specific spectral band close to the ZNW, which in turns causes the soliton to be phase-matched with the same dispersive-wave frequencies all throughout its evolution. Therefore, the intensity of the dispersive wave is enhanced for higher values of $\gamma_1$. On the contrary, the soliton in Fig. \ref{fig:Evos}(c) keeps red-shifting with distance, hence disrupting the phase matching and emitting a much weaker dispersive wave. The emission of a more intense dispersive wave also leads to larger spectral recoil of the soliton [compared to Fig.\ \ref{fig:Evos}(c)]. It is important to emphasize that the input central wavelength in Fig.~\ref{fig:Evos6} is still $1060$~nm, but the DW emission and its associated spectral recoil happen within the first millimeters of propagation. The increased spectral power in the non-solitonic region between $1150$ and $1800$ nm is therefore due to a stronger spectral recoil, as these frequency components are absent in Fig.\ \ref{fig:Evos}(c). In the temporal domain, we see clearly the Raman soliton, a DW, and radiation emitted in the non-solitonic region. SSFS normally causes solitons to decelerate continuously, resulting in a curved trajectory. However, when the SSFS is suppressed after an initial red shift, the soliton keeps its group velocity constant after an initial change. This is why the temporal evolution in Fig.~\ref{fig:Evos6}(b) shows a nearly straight trajectory for the soliton.

\begin{figure}[tb!]
\centering
\includegraphics[width=\linewidth]{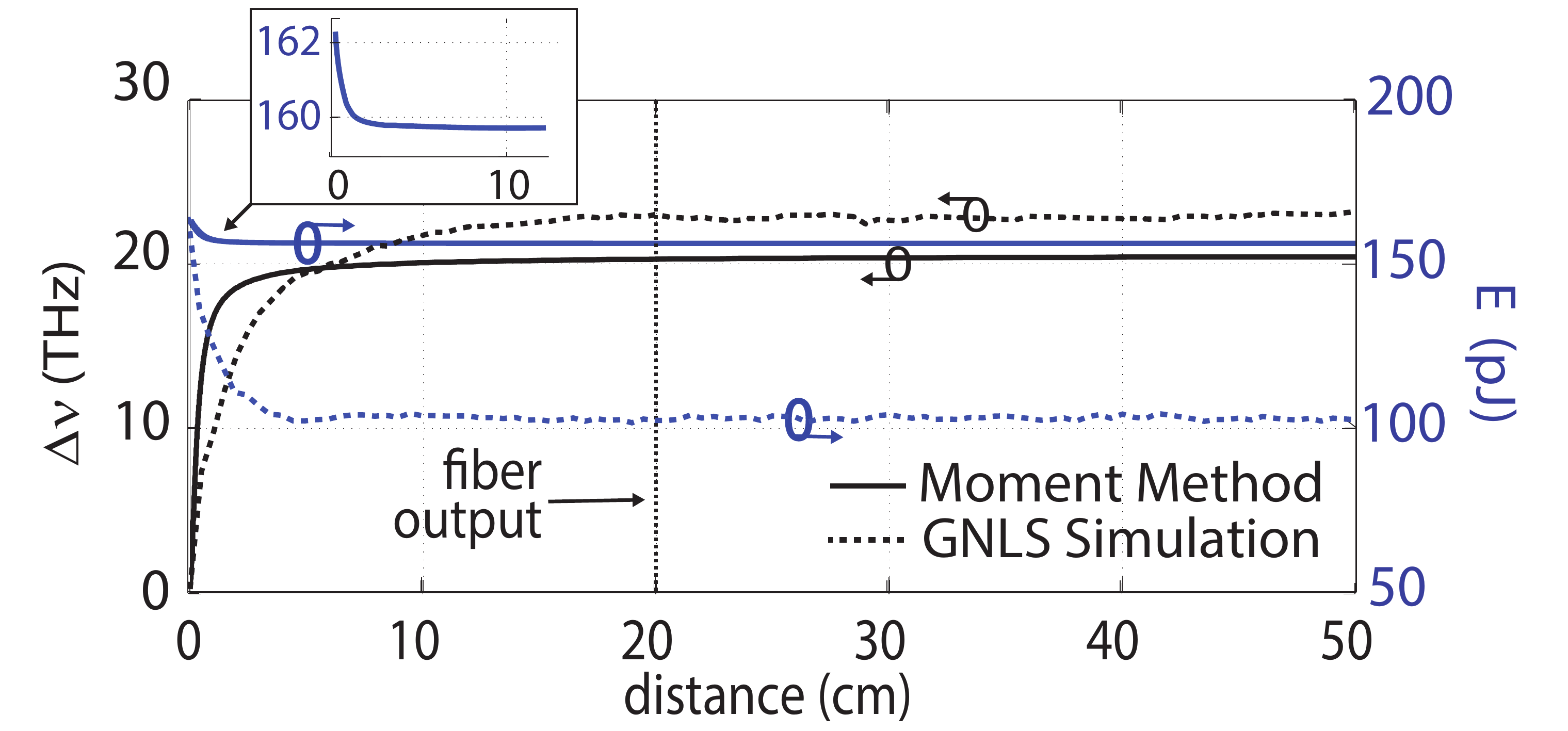}
\caption{(Color online) Moment method predictions for the frequency shift (solid line) and energy (dashed line) over $50$-cm with using the same parameters of Fig. \ref{fig:Evos}, except that $\gamma_1=0.6$. The Vertical dashed line marks the output of evolutions in Fig. \ref{fig:Evos6}.}
\label{fig:energy0p6}
\end{figure}

We briefly discuss the predictions of the moment method for the results shown in Fig.\ \ref{fig:Evos6}. Using Eqs.\ (\ref{E})--(\ref{Omega}), we calculate the frequency shift ($\nu = \Omega/(2\pi)$) and pulse energy $E$ for a $50$~cm long fiber, and the results are shown in Fig.~\ref{fig:energy0p6}. Both of these quantities change rapidly initially but almost stop changing after 10~cm and remain stable after that, indicating that SSFS is being suppressed without additional losses (assuming negligible fiber loss over a short PCF segment used here). Therefore, this mechanism for SSFS suppression does not dissipate the soliton. The frequency shift predicted by the moment method is in decent agreement with the simulation results, but the predicted soliton energy is $50\%$ higher. This discrepancy can be attributed to the moment method, which assumes all of the input energy belongs to the soliton. In numerical simulations, part of the pulse energy is in the non-solitonic region and the soliton rapidly reshapes itself and loses some energy by emitting a blue dispersive wave and also by leaving some pump remnants behind (marked as RW in Fig.\ \ref{fig:Evos6}).

\begin{figure}[tb!]
\centering
\includegraphics[width=8cm]{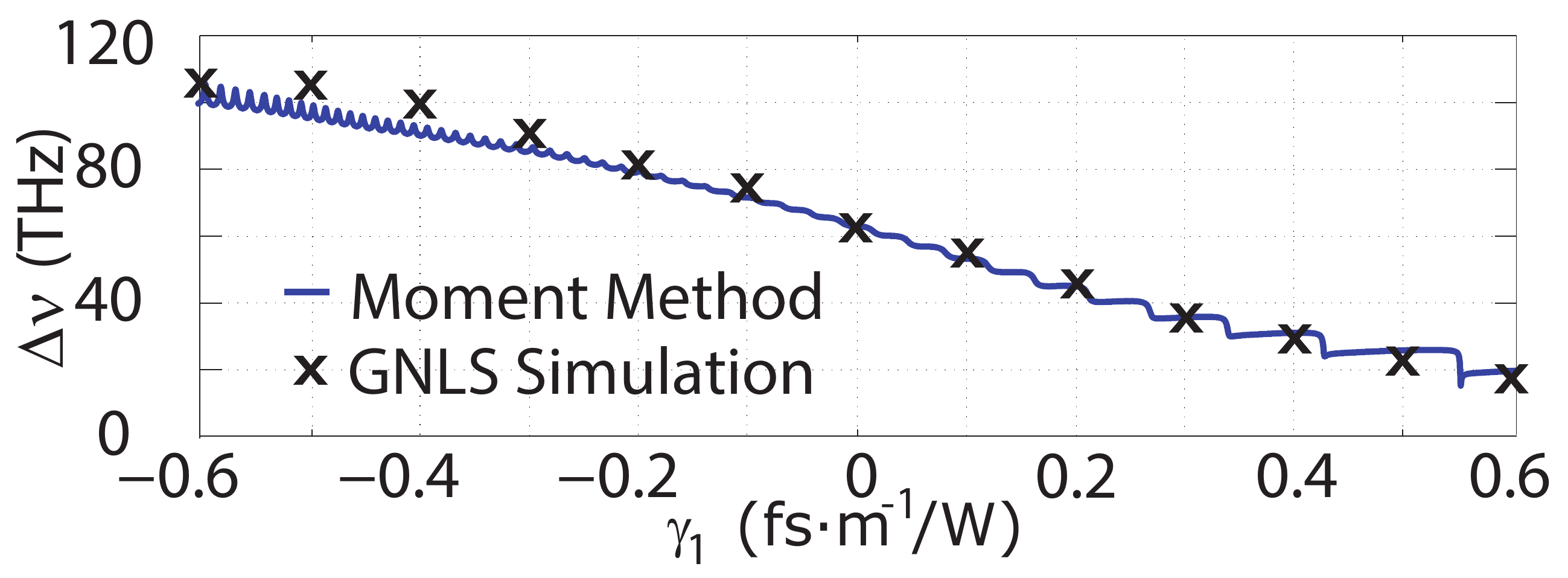}
\caption{(Color online) Frequency shift ($\Delta\nu=|\Omega|/2\pi$) as a function of $\gamma_1$ at the output of the $20$~cm long PCF from the GNLSE simulations (x's) and from solving the moment method Eqs. (\ref{E})-(\ref{Omega}) (dashed line).}
\label{fig:NuVg1}
\end{figure}

Finally, to study how the SSFS depends on $\gamma_1$, we plot in Fig.~\ref{fig:NuVg1} the shift in soliton's central frequency shift ($\Delta\nu=|\Omega|/2\pi$) at the output end of a 20-cm-long PCF as a function of $\gamma_1$ using both full numerical simulations and the moment method. As seen there, the SSFS exceeds 100~THz for $\gamma_1= -0.6$, decreases continually as $\gamma_1$ increases, and reduces to below 20~THz for $\gamma_1= 0.6$. The agreement between full numerical simulations and the moment method is quite reasonable with slight discrepancy for negative values of $\gamma_1$. Again, the likely explanation for the mismatch is that the moment method assumes that all of the input energy belongs to the forming soliton, whereas in simulations a part of the input energy lies in the non-solitonic region.

\section{Conclusions}

To conclude, we studied numerically the propagation of femtosecond pulses in PCFs whose Kerr nonlinearity varies considerably with frequency. Assuming a linear dependence of the nonlinear parameter on frequency, we varied the frequency slope $\gamma_1$ of the nonlinearity over a wide range. Numerical simulations show clearly that the SSFS is enhanced when $\gamma_1$ is negative and is reduced when $\gamma_1$ is positive. For large enough positive values of $\gamma_1$, the zero-nonlinearity wavelength is close to the input wavelength of the pulse on the red side. In this case, the input pulse evolves rapidly toward a fundamental Raman soliton and its energy, soliton order, and central frequency stop changing, leading to a complete suppression of the SSFS without loss of any energy owing to spectral recoil. The predictions of the moment method for the output central frequency were shown to be in very good agreement with numerical simulations.



\begin{thebibliography}{99}

\bibitem{Dudley:06} J. M. Dudley, G. Genty, and S. Coen, Rev. Mod. Phys. \textbf{78}, 1135–1184 (2006).

\bibitem{Skryabin:10} V. Skryabin and A. V. Gorbach, Rev. Mod. Phys. \textbf{82}, 1287–1299 (2010).

\bibitem{Gordon:86} J. P. Gordon, Opt. Lett. {\bf 11,} 662--664 (1986).

\bibitem{Mitschke:86} F. M. Mitschke and L. F. Mollenauer, Opt. Lett. \textbf{11}, 659-661 (1986)

\bibitem{Akhmediev:95} N. Akhmediev and M. Karlsson, Phys. Rev. A \textbf{51}, 2602–2607 (1995).

\bibitem{Roy:11} S. Roy, S. K. Bhadra, and G. P. Agrawal, Curr. Sci. \textbf{100}, 321–342 (2011).

\bibitem{Gorbach:07} A. V. Gorbach and D. V. Skryabin, Nat. Photonics \textbf{1}, 1749–4885 (2007).

\bibitem{Gorbach:07b} A. V. Gorbach and D. V. Skryabin, Phys. Rev. A \textbf{76}, 053803

\bibitem{Driben:13} R. Driben, A. V. Yulin, A. Efimov, and B. A. Malomed, Opt. Express \textbf{21}, 19091-19096 (2013)

\bibitem{Judge:09} A. C. Judge, O. Bang, B. J. Eggleton, B. T. Kuhlmey, E. C. Mägi, R. Pant, and C. Martijn de Sterke, J. Opt. Soc. Am. B \textbf{26}, 2064-2071 (2009)

\bibitem{Pant:10} R. Pant, A. C. Judge, E. C. Magi, B. T. Kuhlmey, M. de Sterke, and B. J. Eggleton, J. Opt. Soc. Am. B \textbf{27}, 1894-1901 (2010)

\bibitem{arteaga:14} F. R. Arteaga-Sierra, C. Mili\'an, I. Torres-G\'omez, M. Torres-Cisneros, G. Molt\'o, and A. Ferrando, Opt. Express \textbf{22}, 23686-23693 (2014).

\bibitem{antikainen:17} A. Antikainen, F. R. Arteaga-Sierra, and G. P. Agrawal, Phys. Rev. A \textbf{95}, 033813 (2017).

\bibitem{Skryabin:03} D. V. Skryabin, F. Luan, J. C. Knight, and P. St. J. Russell, Science 301, 1705–1708, (2003).

\bibitem{Biancalana:04} F. Biancalana, D. V. Skryabin, and A. V. Yulin, Phys. Rev. E \textbf{70}, 016615, (2004).

\bibitem{Bose:16a} S. Bose, R. Chattopadhyay, S. Roy, and S. K. Bhadra, J. Opt. Soc. Am. B \textbf{33}, 1014-1021 (2016).

\bibitem{Bose:16b} S. Bose, A. Sahoo, R. Chattopadhyay, S. Roy, S. K. Bhadra, and G. P. Agrawal, Phys. Rev. A {\bf 94}, 043835, (2016).

\bibitem{Driben:09} R. Driben, A. Husakou, and J. Herrmann, Opt. Express \textbf{17}, 17989-17995 (2009).

\bibitem{Driben:10} R. Driben and J. Herrmann, Opt. Lett. \textbf{35}, 2529-2531 (2010).

\bibitem{Moses:06} J. Moses and F. W. Wise, Phys. Rev. Lett. \textbf{97}, 073903 (2006).

\bibitem{Agrawal:13} G. P. Agrawal, \textit{Nonlinear Fiber Optics}, 5th ed. (Academic, 2013).

\bibitem{Ranka:00} J. K. Ranka, R. S. Windeler, and A. J. Stentz, Opt. Lett. {\bf 25}, 25–27. (2000).

\bibitem{Santhanam:03} J. Santhanam, G. P. Agrawal, Optics Communications, \textbf{222}, 413-420 (2003)

\bibitem{Chen:10} Z. Chen, A. J. Taylor, and A. Efimov,J. Opt. Soc. Am. B \textbf{27}, 1022-1030 (2010)

\end{thebibliography}

\end{document}